\documentstyle[12pt,epsfig]{article}
\newcommand{\mb}{\makebox}
\newcommand{\lt}{\left}
\newcommand{\rt}{\right}

\newcommand{\hf}{\hspace*{\fill}}

\newcommand{\bt}{\begin{tabular}}
\newcommand{\et}{\end{tabular}}
\newcommand{\bq}{\begin{quote}}
\newcommand{\eq}{\end{quote}}
\newcommand{\bd}{
\begin{document}}
\newcommand{\ed}{\end{document}}
\newcommand{\bc}{\begin{center}}
\newcommand{\ec}{\end{center}}
\newcommand{\beqa}{\begin{eqnarray*}}
\newcommand{\eeqa}{\end{eqnarray*}}
\newcommand{\beqn}{\begin{eqnarray}}
\newcommand{\eeqn}{\end{eqnarray}}
\newcommand{\bbibl}{\begin{thebibliography}{99}}
\newcommand{\ebibl}{\end{thebibliography}}
\newcommand{\beq}{\begin{equation}}
\newcommand{\eeq}{\end{equation}}
\newcommand{\ba}{\begin{array}}
\newcommand{\ea}{\end{array}}
\newcommand{\bdm}{\begin{displaymath}}
\newcommand{\edm}{\end{displaymath}}

\newcommand{\R}{\mb{$I\!\!R$}}
\newcommand{\C}{{\cal C}}
\newcommand{\M}{{\cal M}}
\newcommand{\N}{{\cal N}}
\newcommand{\B}{{\cal B}}
\newcommand{\X}{{\cal X}}
\newcommand{\Y}{{\cal Y}}
\newcommand{\F}{{\cal F}}
\newcommand{\Rc}{{\cal R}}
\newcommand{\A}{{\cal A}}
\renewcommand{\P}{{\cal P}}
\renewcommand{\S}{{\cal S}}
\renewcommand{\a}{\alpha}
\renewcommand{\b}{\beta}
\newcommand{\g}{\gamma}
\newcommand{\G}{\Gamma}
\renewcommand{\d}{\delta}
\newcommand{\D}{\Delta}
\newcommand{\th}{\theta}
\newcommand{\e}{\varepsilon}
\newcommand{\eps}{\epsilon}
\newcommand{\h}{\eta}
\renewcommand{\l}{\lambda}
\renewcommand{\L}{\Lambda}
\newcommand{\m}{\mu}
\newcommand{\n}{\nu}
\newcommand{\p}{\pi}
\newcommand{\s}{\sigma}
\newcommand{\Si}{\Sigma}
\newcommand{\ph}{\phi}
\newcommand{\Ph}{\Phi}
\renewcommand{\c}{\chi}
\newcommand{\om}{\omega}
\newcommand{\Om}{\Omega}

\newcommand{\la}{\leftarrow}
\newcommand{\ra}{\rightarrow}
\newcommand{\Ra}{\Rightarrow}
\newcommand{\Lra}{\Leftrightarrow}
\newcommand{\lgra}{\longrightarrow}
\newcommand{\Lgra}{\Longrightarrow}
\newcommand{\lglra}{\longleftrightarrow}
\newcommand{\Lglra}{\Longleftrightarrow}

\newcommand{\ti}{\times}
\newcommand{\what}{\widehat}
\newcommand{\del}{\partial}
\newcommand{\bm}[1]{\mb{\boldmath ${#1}$}}
\newcommand{\ot}{\otimes}
\newcommand{\nn}{\nonumber}
\newcommand{\es}{\emptyset}
\newcommand{\ci}{\subseteq}
\newcommand{\cs}{\supseteq}
\renewcommand{\u}{\cup}
\renewcommand{\i}{\cap}
\newcommand{\bu}{\bigcup}
\newcommand{\bi}{\bigcap}
\newcommand{\tri}{\triangle}
\newcommand{\rec}[1]{\frac{1}{#1}}
\newcommand{\f}{\frac}
\newcommand{\sm}[2]{\sum_{#1}^{#2}}
\newcommand{\ld}{\ldots}
\newcommand{\ov}{\overline}
\newcommand{\ol}[1]{$\bar{\mb{#1}}$}
\newcommand{\un}{\underline}
\newcommand{\iy}{\infty}
\newcommand{\qed}{\hf\rule{2mm}{2mm}}
\newcommand{\wt}{\widetilde}
\newcommand{\ds}{\displaystyle}
\newcommand{\nin}{\not\in}

\newcommand{\alter}[2]{\lt\{ \ba {ll}#1 \\ #2 \ea \rt.}
\newcommand{\alt}[4]{\lt\{ \ba{ll}#1 & \mb{if \,\,}#2 \\ #3 & \mb{if
           \,\,}#4 \ea \rt.}
\newcommand{\altn}[4]{\lt\{ \ba{rl}#1 & \mb{if \,\,}#2 \\ #3 & \mb{if
           \,\,}#4 \ea \rt.}
\newcommand{\alto}[6]{ \lt\{ \ba{ll}#1 & \mb{if \,\,}#2 \\ #3 & \mb{if
           \,\,} #4 \\ #5 & \mb{if \,\,}#6 \ea \rt.}
\newcommand{\altero}[5]{\mb{$\lt\{ \ba {ll}#1 & \mb{if \,\,}#2 \\ #3 &
           \mb{if \,\,} #4 \\ #5 & \mb{otherwise} \ea \rt.$}}

\newcounter{cnt1}
\newcounter{cnt2}
\newcounter{cnt3}
\newcounter{cnt4}
\newcommand{\blr}{\begin{list}{$($\roman{cnt1}$)$} {\usecounter{cnt1}
        \setlength{\topsep}{0pt} \setlength{\itemsep}{0pt}}}
\newcommand{\bla}{\begin{list}{$($\alph{cnt2}$)$} {\usecounter{cnt2}
        \setlength{\topsep}{0pt} \setlength{\itemsep}{0pt}}}
\newcommand{\bln}{\begin{list}{$($\arabic{cnt3}$)$} {\usecounter{cnt3}
        \setlength{\topsep}{0pt} \setlength{\itemsep}{0pt}}}
\newcommand{\ble}{\begin{list}{$\bullet$} {\usecounter{cnt4}
        \setlength{\topsep}{0pt} \setlength{\itemsep}{0pt}}}
\newcommand{\el}{\end{list}}

\newcommand{\no}{\noindent}

\newtheorem{Thm}{Theorem}[section]
\newtheorem{Lem}[Thm]{Lemma}
\newtheorem{Cor}[Thm]{Corollary}
\newtheorem{Prop}[Thm]{Proposition}
\newtheorem{Def}[Thm]{Definition}
\newtheorem{Exm}[Thm]{Example}
\newtheorem{Rem}[Thm]{Remark}
\input epsf.sty

\bd

\title{Toxin-allelopathy among phytoplankton species prevents competitive exclusion}
\author{Shovonlal Roy $\&$ Joydev Chattopadhyay\\
\\
{\normalsize{Agricultural and Ecological Research Unit,}} \\
{\normalsize{Indian Statistical Institute,}}\\
{\normalsize{203, B.T. Road, Kolkata 700108, India}} }
\date{}
\maketitle
\newpage

\bc {\bf Abstract} \ec

\no Toxic or allelopathic compounds liberated by toxin-producing
phytoplankton (TPP) acts as a strong mediator in plankton dynamics.
On an analysis of a set of phytoplankton biomass-data that have been
collected by our group in the North-West part of the Bay of Bengal,
and by analysis of a three-component mathematical model under a
constant as well as a stochastic environment, we explore the role of
toxin-allelopathy in determining the dynamic behaviour of the
competing-phytoplankton species. The overall results, based on
analytical and numerical wings, demonstrate that toxin-allelopathy
due to the toxin-producing phytoplankton (TPP) promotes a stable
coexistence of those competitive phytoplankton that would otherwise
exhibit competitive exclusion of the weak species. Our study
suggests that TPP might be a potential candidate for maintaining the
coexistence and diversity of competing phytoplankton species.
\\
\\
\\
\no {\bf Key words:} Phytoplankton, toxin, allelopathy,
coexistence, competitive exclusion, paradox of plankton
\\
\\
\\
\newpage
\section{Introduction}
\no The principle of competitive exclusion ensures that the number
of competing species cannot exceed the number of distinct resources
(Hardin, 1960). Simple competition models and competition
experiments in laboratory also suggest that the number of species
that co-exist in equilibrium can be greater than the number of
limiting resources only if additional mechanisms are involved
(Tilman, 1977, 1981; Somer 1985, 1986; Rothhaupt 1988, 1996;
Scheffer \emph{et al.}, 1997; Huisman and Weissing, 1999).  For
instance, temporal variation in the supply of a single resource may
allow the coexistence of two species (Stewart $\&$ Levin, 1973;
Levins, 1979; Armstrong $\&$ McGehee, 1980). For two competing prey
or parasites, predator or parasite-mediated coexistence is possible
provided that the inferior competitor is resistant to exploitation
(Levin, 1970; Levin \emph{et al}., 1977). Sometimes interference
competition also promotes stable coexistence of two species on a
single resource (e.g., Vance, 1985). Furthermore, in homogeneous
environment inhibitory substances such as pesticides, derived from
external sources can promote stable coexistence of two species
competing for a single resource (Lenski $\&$ Hattingh, 1986). Unlike
the above biological situations, in view of the competitive
exclusion principle the coexistence of a large number of
phytoplankton species on a seemingly limited variety of resources in
natural waters is remarkable; this is referred to as `the paradox of
the plankton' (Hutchinson, 1961). To explain this paradox, several
attempts have been made. Hutchinson (1961) proposed that because of
weather-driven fluctuations, plankton communities are not in
equilibrium. Authors such as Richerson \emph{et al}. (1970) argued
in a fashion similar to Hutchinson (1961) that continuous variation
in environmental conditions, due to seasonal cycles and less
predictable factors such as weather, offer the most likely solution.
On the other hand, theoretical studies predict that competition
among different species of phytoplankton can generate oscillations
and chaos, which may in turn promote their coexistence (Huisman $\&$
Weissing, 1999). However, none of these explanations is universally
accepted.

In an aquatic ecosystem, some species of plankton liberate ``toxic"
or ``allelopathic agents" that affect the growth of other
micro-algae (Hallam \emph{et al}, 1983; Arzul \emph{et al} 1999).
Among marine algae, allelopathy was observed  both \emph{in vitro}
and \emph{in situ} (e.g Chan \emph{et al}., 1980; Nielsen \emph{et
al}., 1990; Schmidt and Hansen, 2001, Tillmann and John, 2002;
Fistarol \emph{et al}., 2003, 2004), however, the chemical nature
and role of allelopathic compounds remained poorly understood (Sole
\emph{et al}., 2005). In phytoplankton-zooplankton interactions,
toxicity acts as a strong mediator (Kozlowski-Suzuki \emph{et al.},
2003). Efforts have been made to study the role of toxin inhibition
on zooplankton (e.g. Chattopadhyay \emph{et al.}, 2002a, 2002b;
Sarkar $\&$ Chattopadhyay, 2003). Recently, Roy \emph{et al}.
reported that in regulating non-equilibria of a
phytoplankton-zooplankton system, toxin inhibition on zooplankton
caused by toxin-producing phytoplankton (TPP) acts as a driving
force. However, allelopathic interaction among the phytoplankton
species has not been included in that study. Among the algae
species, toxin-allelopathy is an important chemical-signaling
process (see, review by Cembella, 2003). Interactions between two
allelopathic-species was studied mathematically by many authors
(e.g. Maynard-Smith, 1974; Chattopadhyay, 1996; Mukhopadhyay
\emph{et al.}; 1998; Nakamaru $\&$ Iwasa, 2000). Schimdt and Hansen
(2001) made a laboratory experiment on plankton allelopathy in which
15 species of marine phytoplankton were exposed to suspensions of a
toxic alga known as \emph{Crysocomulina polylepis}. Recently, Sole
\emph{et al}. (2005) used those experimental data to estimate the
allelopathic parameters based on a model proposed by Chattopadhyay
(1996). The study of Sol\'{e} \emph{et al}. (2005) suggests a
functional form suitable for quantifying the strength of
allelopathic interaction between toxic and non-toxic algae.

\par However, in the previous studies little attention has been paid
to explore the role of allelopathic interaction on the coexistence
and persistence of phytoplankton species competing for the same
resources. The objective of this article is to investigate the role
of toxin-allelopathy in maintaining the coexistence of the
competitive-phytoplankton species in the marine ecosystem. On
analysis of a set of field-data that we have collected from the
North-West coast of the Bay of Bengal, we propose that a possible
role of toxic phytoplankton might be responsible for a stable
coexistence of the competing phytoplankton. Next we formulate a
simple three-component model for describing the interaction among
two non-toxic phytoplankton and a toxic phytoplankton. We analyze
the model in a deterministic and a stochastic environment, and find
suitable bounds on the allelopathic parameters under which a stable
coexistence of the competing species is possible. Through numerical
experiments, we support our analytical findings and demonstrate the
role of toxin allelopathy in maintaining the stable coexistence of
those competing phytoplankton that would otherwise exhibit an
exclusion of the weak species. The study demonstrates that
toxin-allelopathy among phytoplankton species counteracts
competitive exclusion.

\par The organization of the paper is as follows. Section 2 produces
a qualitative analysis of the plankton dynamics based on the field
observation on non-toxic and toxic phytoplankton. Section 3 proposes
a three-component mathematical-model consisting of
two-competitive-phytoplankton and a toxic phytoplankton. The model
is analyzed to find the criterion for coexistence and persistence of
the species. In Section 4, by incorporating stochastic perturbation,
the dynamic behaviour is studied under environmental fluctuations.
In Section 5 we present numerical experiments to support the
analytical results. We discuss the overall results of our study in
Section 6.
\section{Field Observation}
\no Since 1999, the monitoring and identification (Tomas, 1997) of
marine plankton population has been carried out by our group in the
North - West coast of the Bay of Bengal (for detail see,
Chattopadhyay \emph{et al}., 2002). A significant number of species
of phytoplankton have been identified that produce toxic or
inhibitory compounds (Chattopadhyay \emph{et al.}, 2002a, 2002b;
Sarkar $\&$ Chattopadhyay, 2003). The toxin-producing phytoplankton
(TPP) group contains (i) planktonic or benthic micro-algae that
produce toxin (e.g., the motile stage of \textsl{Alexandrium}, the
benthic \textsl{Gambierdiscus}), (ii) other toxic dinoflagellates
(e.g., \textsl{Pfiesteria}), (iii) macroalgae that results in
noxious smells (e.g. \textsl{Pilayella}), (iv) a few species of
Cyanobacteria or blue algae (e.g., \textsl{Microcystis}), (v)
non-toxic microorganisms that result in hypoxic conditions (e.g.
\textsl{Chaetoceros, Mesodinium}). For a detailed list of TPP
species identified by our group, see Chattopadhyay \emph{et al.},
2002.

\par For understanding the interaction between non-toxic and toxic algae
at species level, we choose from the list of phytoplankton species
that have been identified during the period 2000-2001, a combination
three species consisting of two non-toxic and a toxic phytoplankton.
We choose those algae that were present at significant biomass
throughout the study period. The three species chosen belong to
diatom group. The two non-toxic phytoplankton (NTP) are
\emph{Coscinodiscus sp} (say species 1, biomass at any time $x_1$)
and \emph{Biddulphia sp} (species 2, biomass $x_2$); and the
toxin-producing phytoplankton (TPP, say species 3, biomass $x_3$) is
\emph{Chaetoceros sp} (as cited by Chattopadhyay \emph{et al.},
2002). The abundance level of all the species are fluctuating over
the time (Figure 1), and throughout the period of observation the
abundance level of \emph{Coscinodiscus sp} is higher
\begin{figure}
\epsfxsize =6.0in \epsfysize = 6.5in \epsfbox{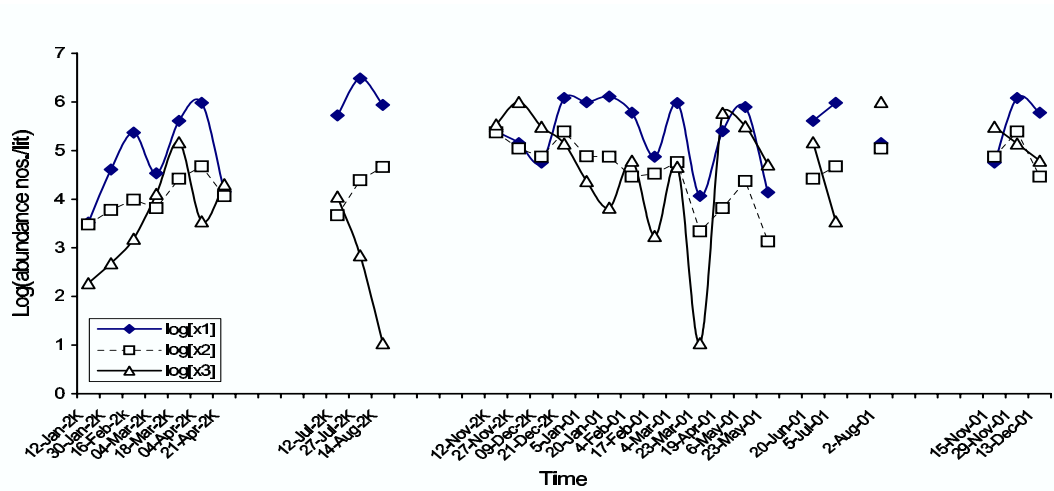}
\no\caption{Coexistence of two non-toxic competitive-phytoplankton
species with a toxic species. Here, $x_1$ is the biomass at any time
point of the NTP (species 1) \emph{Coscinodiscus sp}, $x_2$ is that
of NTP (species 2) \emph{Biddulphia sp}, and $x_3$ is the same of
TPP (species 3) \emph{Chaetoceros sp}; \emph{Coscinodiscus sp} is
the most abundant species and stronger competitor than
\emph{Biddulphia sp}. The gaps in the axis of collection represents
the time point when the sampling was suspended due to several
reasons.}
\end{figure}
than that of \emph{Biddulphia sp}. Now, because all the species
interact in a common marine environment, in principle they compete
for the common resources available (such as sunlight, dissolved
nutrient). In our study region, among the chosen species
\emph{Coscinodiscus sp} is the most dominant in biomass throughout
the sampling period (see Figure 1). Because the species have been
identified in a common sampling and from a common field, it is
reasonable to assume that the ecological and biological factors that
affect the growth of species are similar for all the species.
Moreover, in a general sense a potential role of competitive effect
of one species would be to hamper the abundance level of the other
species. So, due to the lack of any other experimental data, we may
consider the abundances of two non-toxic species (where
toxin-allelopathy does not come into play) as a potential indicator
of the dominance level of resource competition. Clearly this
argument does not hold for a toxic and a non-toxic species. In this
sense \emph{Coscinodiscus sp} is a stronger competitor than
\emph{Biddulphia sp}. The distribution of the abundance ratio of
\emph{Coscinodiscus} to \emph{Biddulphia}, when plotted against the
abundance of the toxic phytoplankton, depicts a decreasing trend for
higher biomass of toxic phytoplankton (Figure 2). Pearson
correlations confirm this trend. The correlation coefficient between
the abundance of TPP (in log scale) and the abundance ratio of $x_1$
to $x_2$ is ($r=-0.515$), which is significant at 5$\%$ level. On
the other hand, the total biomass of the two non-toxic algae has a
significant positive correlation with the $x_1/x_2$ coefficient
($r=0.40$, $P<0.05$). However, if we include the toxic algae the sum
of the biomass of all the species has insignificant correlation
($r=0.24$, $P>0.1$) with the $x_1/x_2$ coefficient. These results
suggest that, when the overall biomass of two non-toxic algae (that
influences the resource competition) increases, the abundance ratio
of the two algae also increases significantly; consequently in the
system the pressure of \emph{Coscinodiscus} is enhanced and that of
the \emph{Biddulphia sp} is reduced. However, the scenario changes
significantly if the presence of a toxic algae is taken into
consideration. The abundance ratio $x_1/x_2$ shows a significantly
reverse trend (Figure 2).
\begin{figure}
\epsfxsize = 5.0in \epsfysize = 6in \epsfbox{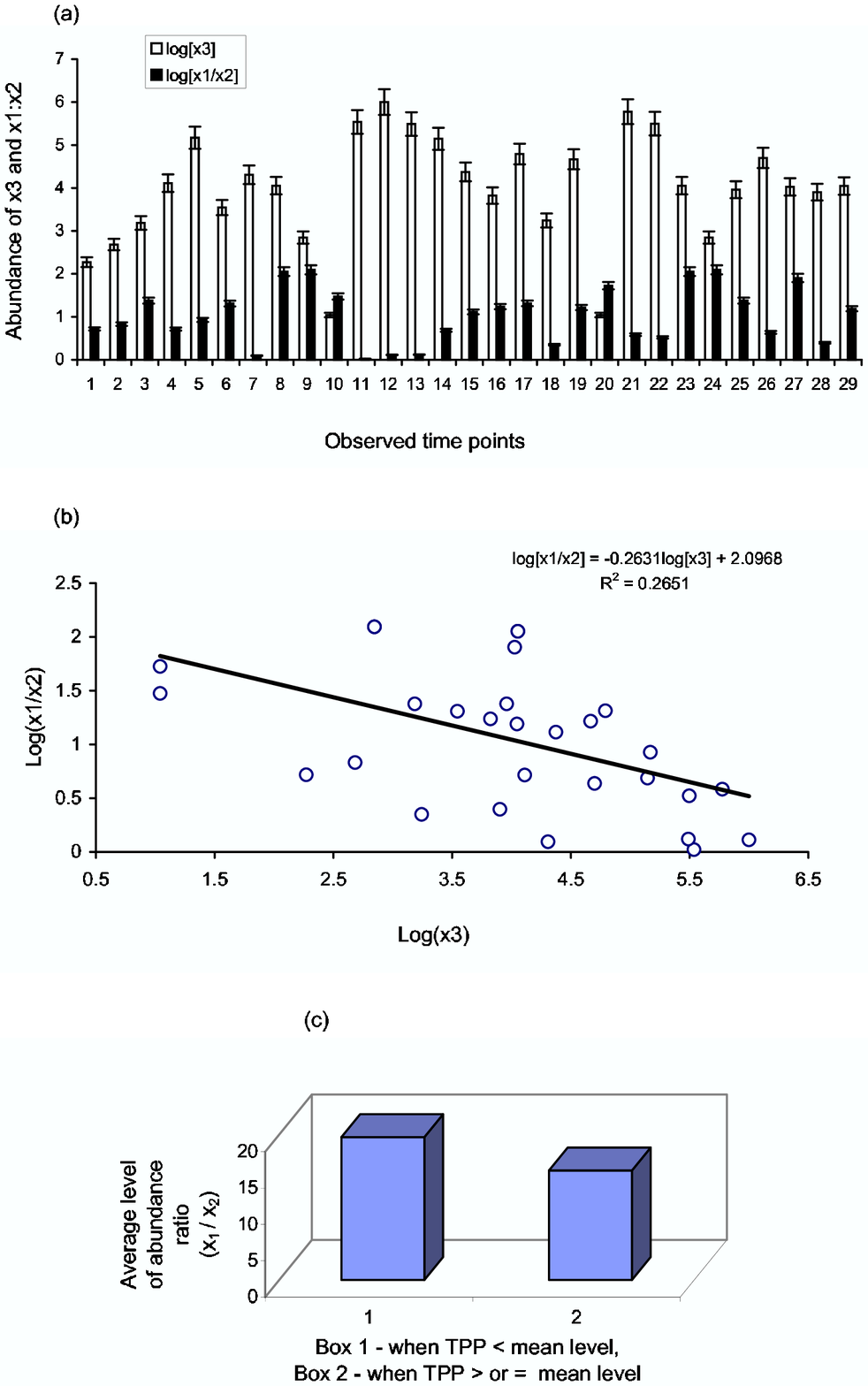}
\no\caption{Distribution of the abundance ratio of
\emph{Coscinodiscus sp} to \emph{Biddulphia sp} and the abundance of
\emph{Chaetoceros sp}. 2(a) Bar diagram depicting the abundance
ratio of $x_1$ to $x_2$  and  abundance of $x_3$. Log($x_1$/$x_2$)
reduces by 25 $\%$ when TPP is sufficiently present in the system.
2(b) Abundance ratio of \emph{Coscinodiscus sp} to \emph{Biddulphia
sp} against the abundance of \emph{Chaetoceros sp} (in log form)
depicts a negative slope of the trend line. 2(c) Levels of average
abundance ratios $x_1/x_2$ corresponding to TPP biomass at `less
than mean level' ($M_1$) and `greater than mean level' ($M_2$).
Compared with category $M_1$, the average level of $x_1/x_2$
decreases to category $M_2$ by 25$\%$}
\end{figure}
\par To compare the ratio $x_1$ to $x_2$ in situations of high and low
TPP abundance, we divide the TPP biomass in two categories: `less
than mean value' (say $M_1$) i.e., when TPP abundance is less than
its overall average over the observed time points, and `greater than
or equal to mean value' (say $M_2$) i.e., when TPP abundance is
greater than or equal to its overall average over the observed time
points. We say TPP is at low abundance at any time if the TPP
biomass belongs to category $M_1$, and that TPP is at high abundance
if TPP biomass belongs to $M_2$. Now, corresponding to those time
points where TPP biomass belongs to $M_1$, let the average of the
$x_1/x_2$ ratios be $m_1$. And corresponding to category $M_2$, let
the average of the $x_1/x_2$ ratios be $m_2$. The the quantity
$\delta_m\,=\,\f{(m_2-m_1)}{m_1}\times 100$ may represent the change
in the mean level of the abundance of \emph{Coscinodiscus} to
\emph{Biddulphia} when the biomass of \emph{Chaetoceros sp} is
sufficiently high compared with when its abundance is low in the
system. We find that $\delta_m$ is around (-25$\%$), i.e., the mean
level of $x_1/x_2$ reduces by 25$\%$ when TPP abundance is high
(see, Figure 2(c)). These results suggest that the presence of toxic
phytoplankton is favourable for the existence of the weak species in
marine ecosystem. Hulot $\&$ Huisman (2004) claimed that, because of
the toxic compounds released by the TPP, the competitive
disadvantage between phytoplankton species is reduced. Our field
observation also resembles the claim. By releasing allelopathic
chemicals, the toxic species of phytoplankton gain a significant
advantage in resource competition. Moreover, these toxic chemicals
affect significantly the growths of the other competitors. Due to
the presence of toxic chemicals, the species of non-toxic
phytoplankton can hardly impose any competitive effect on these
allelopathic species. So, the competition coefficient between a
toxic and a non-toxic phytoplankton is negligible (Sol\'{e} et al.
2005). In this way, in a mixed-species environment, the allelopathic
species exhibit a passive mutualism towards the weak species, and
promote those species to survive in competition (also found in Roy
\emph{et al.} \emph{submitted}). Although not presented here, some
other triad of two non-toxic and a toxic species that are present in
dominant biomass would also exhibit a similar dynamics (Roy \emph{et
al.} submitted). We would like to mention that while using the
correlation analysis and linear regression, we have ignored the data
autocorrelation, a well-known analysis for a time series. However,
because the data series considered is short and discontinuous, it is
difficult to use the techniques specific for time series analysis.
An entirely different approach for estimation of the missing values
by an imputation method called Expectation-Maximization, and
analysis of autocorrelation by a Vector Auto-regressive model also
supports the results obtained here; the details of this analysis is
reported in Roy \emph{et al.} \emph{submitted}.

\par Based on these arguments, to explore and display the dynamic behaviour of the competing
phytoplankton species taking into account the presence of
allelopathic species, in the following sections we propose and
analyze a simple mathematical model. The main objective of the
analysis of the model is to find suitable mathematical bounds on the
toxin-allelopathy parameters, under a constant environment as well
as under a stochastic environment.
\section{The Mathematical Model}
\no To develop a mathematical model for describing the interaction
among two non-toxic phytoplankton (species 1 with biomass $x_1$ and
species 2 with biomass $x_2$) and a toxic phytoplankton (species 3
with biomass $x_3$), we make the following assumptions,
\\
\no (i) The non-toxic phytoplankton species (species 1 and 2)
compete for the same resource following the Lotka-Volterra
competition model, where species 1 is the stronger competitor than
species 2.
\\
\no (ii) Allelopathic interactions between a non-toxic and a
toxin-producing phytoplankton is described by a nonlinear function
suggested by Sol\'{e} \emph{et al.} (2005).
\\
\no (iii) Competitive interaction between a non-toxic and a toxic
phytoplankton is negligible (Sol\'{e} \emph{et al. }, 2005)
\par Based on the above assumptions, the interaction among two
non-toxic and a toxic phytoplankton is represented in the following
mathematical model:
\begin{eqnarray}
\f{dx_1}{dt} &=& {\it x_1}\,\left ({\it r_1}-{\it \alpha_1}\,{\it
x_1}-{\it \beta_{12}}\,{\it x_2 }-{\it \gamma_1}\,{\it x_1}\,{{\it
x_3}}^{2}\right ), \nonumber
\\
\frac{dx_2}{dt}&=&{\it x_2}\,\left ({\it r_2}-{\it \alpha_2}\,{\it
x_2}-{\it \beta_{21}}\,{\it x_1 }-{\it \gamma_2}\,{\it x_2}\,{{\it
x_3}}^{2}\right ),
\\
\frac{{\it dx_3}}{dt} &=& {\it x_3}\,\left ({\it r_3}-{\it
\alpha_3}\,{\it x_3}\right ), \nonumber
\end{eqnarray}
\no The model is analyzed under the following initial conditions:
\begin{eqnarray}
x_1(0) >\,0,~ x_2(0) \,>\, 0,~ x_3(0)\,>\, 0.
\end{eqnarray}
\no Here, $r_i$ ($i=1,2,3$) are the specific growth rates of species
$i$, $\a_i$ are the coefficients of intraspecific competition,
$\beta_{12}$ and $\beta_{21}$ are the interspecific competition
coefficients between $x_1$ and $x_2$, $\gamma_i$ ($i=1,2$) are the
strengths of toxin-allelopathy between toxic and non-toxic
phytoplankton.

\subsection{Local stability analysis}
\no The model system (1) has the following equilibria:
\\
~$E_0 \left(0,\,0,\,0 \right)$, $~~~E_1\left ({\frac{{\it
r_1}}{{\it\a_1}}},\,0,\,0 \right ),~$ $~~E_2\left
(0,\,\f{r_2}{\a_2},\,0\right),~~$ $~~E_3
\left(0,\,0,\,\f{r_3}{\a_3}\right),\\ ~~$ $E_4 \left
({\frac{\a_2\,r_1\,-\,\beta_{12}\,r_2}{\a_1\,\a_2\,-\,\beta_{12}\,\beta_{21}}},
\,\f{\a_1\,r_2\,-\,\beta_{21}\,r_1}{\a_1\,\a_2\,-\,\beta_{12}\,\beta_{21}},\,0
\right),~~~$ $~~E_5\left
(\f{r_1\,\a_3^2}{\a_1\,\a_3^2\,+\,\g_1\,r_3^2},\,0,\,\f{r_3}{\a_3}
\right),~~$
$~~E_6\left(\,0,\,\f{r_2\,\a_3^2}{\a_2\,\a_3^2\,+\,\g_2\,r_3^2},\,\f{r_3}{\a_3}\right)~~$,
and the interior equilibrium $E^*(\,x_1^*,\,x_2^*,\,x_3^*)$,

\no where,
\begin{eqnarray}
x_1^*\,&=&\,{\frac {{{\it \alpha_3}}^{2}{{\it r_3}}^{2}{\it
r_1}\,{\it \gamma_2}+{{\it \alpha_3}}^{4}\left ({\it \alpha_2}\,{\it
r_1}\,-{\it r_2}\,{\it \beta_{12}} \right )}{\left ({\it
\alpha_2}\,{{\it \alpha_3}}^{2}+{\it \gamma_2}\,{{\it
r_3}}^{2}\right )\,{{\it r_3}}^{2}\,{\it \gamma_1}+{{\it
\alpha_3}}^{2}\left (- {{\it \alpha_3}}^{2}{\it \beta_{21}}\,{\it
\beta_{12}}+{{\it \alpha_3}}^{2}{\it \alpha_2}\,{\it \alpha_1}+{\it
\gamma_2}\,{{\it r_3}}^{2}{\it \alpha_1}\right)
}},\\
{x_2}^*\,&=&\,{\frac{{{\it \alpha_3}}^{2}{{\it r_3}}^{2}{\it
r_2}\,{\it \gamma_1}+{{\it \alpha_3}}^{4}\left ({\it r_2}\, {\it
\alpha_1}-{\it r_1}\,{\it \beta_{21}}\right ) }{\left ({\it
\alpha_2}\,{{\it \alpha_3}}^{2}+{\it \gamma_2}\,{{\it
r_3}}^{2}\right )\,{{\it r_3}}^{2}\,{\it \gamma_1}+{{\it
\alpha_3}}^{2}\left (- {{\it \alpha_3}}^{2}{\it \beta_{21}}\,{\it
\beta_{12}}+{{\it \alpha_3}}^{2}{\it \alpha_2}\,{\it \alpha_1}+{\it
\gamma_2}\,{{\it r_3}}^{2}{\it \alpha_1}\right)}}
\end{eqnarray}
and
\begin{eqnarray}
x_3^*\,=\,{\frac {{\it r_3}}{{\it \alpha_3}}}.
\end{eqnarray}
\no For any non-negative set of values of the model-parameters, the
equilibria $E_0$, $E_1$, $E_2$, $E_3$, $E_5$ and $E_6$ exist. A
sufficient condition on the parameters for feasibility of $E_4$ is
\begin{eqnarray}
\a_1\,\a_2\,&>&\,\beta_{12}\,\beta_{21}
\end{eqnarray}
\no The interior equilibrium $E^*$ exists if the following set of
inequalities hold
\begin{eqnarray}
\gamma_1\,&>&\,{\frac{{\a_3}^2\,r_1}{{r_3}^2\,r_2}}\,\beta_{21},
\\
\gamma_2\,&>&\,\max\,\{~{\frac{{\a_3}^2\,r_2}{{r_3}^2\,r_1}}\,\beta_{12},\,
{\frac{{\a_3}^2\,\left(\beta_{12}\,\beta_{21}\,-\,\a_1\,\a_2\right)}{\a_1\,{r_3}^2}}~\}.
\end{eqnarray}
\no We find that the coexistence of the interior equilibrium depends
on the strength of the toxin-allelopathy parameters. On generating
the community matrix, we perform local-stability analysis (LAS) of
the model system (1) around each biologically feasible equilibrium.
In the following theorem, we summarize the results of the LAS.
\\
\no \textbf{Theorem 3.1:} {\it The boundary equilibria (E$_0$,
E$_1$, E$_2$, E$_3$, E$_4$, E$_5$ and E$_6$) are repellers under the
following conditions,
\begin{eqnarray}
\a_1\,+\g_1\,\left (\f{r_3}{\a_3}
\right)^2\,&>&\,\beta_{21}\,\,{\frac{r_{1}}{r_2}},\\
\a_2\,+\g_2\,\left (\f{r_3}{\a_3}
\right)^2\,&>&\,\beta_{12}\,\,\f{r_2}{r_1}.
\end{eqnarray}
\no The interior equilibrium point $E^*$ is locally asymptotically
stable if the following condition holds}
\begin{eqnarray}
\g_1\,\g_2\,\geq\,\left({\frac{\a_3}{r_3}}\right)^4\,\beta_{12}\,\beta_{21}\,.
\end{eqnarray}
\emph{(Proof is obvious)}
\\
\par In the absence of the toxic phytoplankton ($x_3$), the model
system (1) reduces to the well known Lotka-Volterra (LV)
competition model. It is well established that LV model exhibits
competitive exclusion of one or both the competitors if any one or
both of the following conditions hold (for detail see, Kot, 2001)
\begin{eqnarray}
\beta_{21}\,\f{r_1}{r_2}\,&>&\,\alpha_1\\
\beta_{12}\,\f{r_2}{r_1}\,&>&\,\alpha_2
\end{eqnarray}
\no Now we are in a position to compare the inequalities obtained in
(9)-(10) and (12)-(13). It follows from Theorem (3.1) that, even if
any one or both of the conditions (12)-(13), that are necessary for
competitive exclusion in LV model, is satisfied, toxin allelopathy
due to TPP, the strength of which satisfies the conditions (9)-(10),
promotes coexistence of the competitive phytoplankton species.

\par In the following section, to study the dynamics of the interacting species under a variable
environment, we extend the scope of deterministic model to a
stochastic set up.

\section{The Stochastic Model}

\no We assume that the stochastic perturbations of the variables
around $E^*$ are of white-noise type proportional to the distances
of $x_1,\, x_2,\, x_3$ from the values ${x_1}^*, {x_2}^*,x_3^*$
(Beretta \emph{et al}., 1998). Under this assumption, system (1)
takes the following form

\begin{eqnarray}
dx_1 &=& [\,{\it x_1}\,\left ({\it r_1}-{\it \alpha_1}\,{\it
x_1}-{\it \beta_{12}}\,{\it x_2 }-{\it \gamma_1}\,{\it x_1}\,{{\it
x_3}}^{2}\right )\,]\,dt + \sigma_1(x_1-{x_1}^*)\,d{\xi_t}^{1},
\nonumber
\\
dx_2 &=& [\,{\it x_2}\,\left ({\it r_2}-{\it \alpha_2}\,{\it
x_2}-{\it \beta_{21}}\,{\it x_1 }-{\it \gamma_2}\,{\it x_2}\,{{\it
x_3}}^{2}\right )\,]\,dt + \sigma_2(x_2-{x_2}^*)\,d{\xi_t}^{2},
\\
dx_3 &=& [\,{\it x_3}\,\left ({\it r_3}-{\it \alpha_3}\,{\it
x_3}\right )\,]\,dt + \sigma_3(x_3-x_3^*)\,d{\xi_t}^{3}. \nonumber
\end{eqnarray}

\no Here $\sigma_{i}, ~(i=1,2,3)$ are real constants defined as the
intensities of the stochasticity, and ${\xi_t}^{i} = \xi_{i}(t),
~(i=1,2,3)$ are independent standard Wiener-process (Gikhman and
Skorokhod, 1974, 1975, 1979).

To investigate the robustness of the dynamical behaviour of model
(1), stochastic stability of the interior equilibrium $E^*$ is
studied using the model (14). System (14) can be represented as an
Ito Stochastic Differential Equation of the following type
\begin{eqnarray}
dX_t &=& f(t,X_t)\,dt + g(t,X_t)\,d\xi_t, \nonumber \\
X_{t_0} &=& X_0,~t \in [t_0,t_f],
\end{eqnarray}
\no where the solution $\{X_t, t \in [t_0,t_f]~ (t>0)\}$ is an Ito
process, $f$ is the slowly varying continuous component or drift
coefficient and $g$ is the rapidly varying continuous random
component or diffusion coefficient (Kloeden and Platen, 1995),
$\xi_t$ is a multi-dimensional stochastic process having scalar
Wiener-process components with increments
$\Delta{\xi_t}^{j}={\xi_{t+\Delta t}}^{j}-{\xi_t}^{j}=
\xi_j(t+\Delta t) - \xi_j(t),$ which are independent Gaussian
random-variables $N(0,\Delta t)$.

\no Comparing (14) and (15), we have

\begin{eqnarray} X_t = (x_1,x_2,x_3)^{T},~\xi_t=( {\xi_t}^1, {\xi_t}^2,
{\xi_t}^3)^{T},
\end{eqnarray}
\begin{eqnarray}
 \emph{f} = \left(%
\begin{array}{c}
{\it x_1}\,\left ({\it r_1}-{\it \alpha_1}\,{\it x_1}-{\it
\beta_{12}}\,{\it x_2 }-{\it \gamma_1}\,{\it x_1}\,{{\it
x_3}}^{2}\right ) \\
{\it x_2}\,\left ({\it r_2}-{\it \alpha_2}\,{\it x_2}-{\it
\beta_{21}}\,{\it x_1 }-{\it \gamma_2}\,{\it x_2}\,{{\it
x_3}}^{2}\right ) \\
{\it x_3}\,\left ({\it r_3}-{\it \alpha_3}\,{\it
x_3}\right ) \\
\end{array}%
\right)
\end{eqnarray}
and
\begin{eqnarray}
\emph{g}\,=\,
\left(%
\begin{array}{ccc}
  \sigma_1(x_1-{x_1}^*) & 0 & 0 \\
  0 & \sigma_2(x_2-{x_2}^*)  & 0 \\
  0 & 0 & \sigma_3(x_3-x_3^*) \\
\end{array}%
\right)
\end{eqnarray}

\no Since the diffusion matrix (18) depends on the solution $ X_t
= (x_1,x_2,x_3)^T$, system (14) is said to have multiplicative
noise. From the diagonal form of the diffusion matrix (18), system
(14) is said to have (multiplicative) diagonal noise.

\subsection{Stochastic stability of the interior equilibrium}

\no By defining the variables
$u_1=x_1-{x_1}^*,~u_2=x_2-{x_2}^*,~u_3=x_3-x_3^*$, the stochastic
differential equations (14) can be centered at the interior
equilibrium $E_*$.

\no To show that system (14) is asymptotically stable in mean
square sense (or in probability) we linearize the vector function
$f$ around the positive equilibrium $E^*$. The linearized
stochastic differential equations around $E^*$ (using the
variational matrix $J$) take the following form

\begin{eqnarray} du(t)= f(u(t))\,dt + g(u(t))\,d\xi(t)
\end{eqnarray}

\no where $u(t)=\left(u_1(t),u_2(t),u_3(t)\right)^T$ and

\begin{eqnarray}
f(u(t))\,=\, \left [\begin {array}{c} {\it x_1^*}\,\left (-{\it
\alpha_1}-{\it \gamma_1}\, {{\it x_3^*}}^{2}\right ){\it u_1}-{\it
x_1^*}\,{\it \beta_{12}}\,{\it u_2}-2\,{{ \it x_1^*}}^{2}{\it
\gamma_1}\,{\it x_3^*}\,{\it u_3}\\\noalign{\medskip}-{\it
x_2^*}\,{\it \beta_{21}}\,{\it u_1}+{\it x_2}\,\left (-{\it
\alpha_2}-{\it \gamma_2 }\,{{\it x_3^*}}^{2}\right ){\it
u_2}-2\,{{\it x_2^*}}^{2}{\it \gamma_2}\,{\it x_3^*}\,{\it u_3}\\
\noalign{\medskip}-{\it \alpha_3}\,{\it x_3^*}\,{\it u_3}
\end {array}\right]
\end{eqnarray}

\begin{eqnarray}
g(u(t))\,=\,
\left(%
\begin{array}{ccc}
\sigma_1\,u_1 & 0 & 0 \\
0 & \sigma_2\,u_2 & 0 \\
0 & 0 & \sigma_3\,u_3\\
\end{array}%
\right) ,
\end{eqnarray}

\no with the parametric conditions for existence of $E^*$ stated in
section (3.1). Clearly, the positive equilibrium $E^*$ in equation
(19) corresponds to the trivial solution $(u_1,u_2,u_3)=(0,0,0)$.

\no Let us define a set $\Psi= \{(t \geq t_0) \times R^3,~t_0 \in
R^{+}\}$. Now there exists a function $V \in {C_2^0}(\Psi)$ such
that $V$ is twice continuously differentiable (i.e., a $C^2$
function) with respect to $u$ and continuous (i.e., $C^0$) with
respect to $t$. With reference to (19), we define the following
function:
\begin{eqnarray}
W(u,t) = \f{\partial V(u(t),t)}{\partial t} + f^T (
u(t))\,\f{\partial V(u,t)}{\partial u} +
\f{1}{2}\,Tr\,[\,g^T(u(t))\,\,\f{\partial^2V(u,t)}{\partial
u^2}\,\,g(u(t))\,],
\end{eqnarray}
\no where $\f{\partial V}{\partial u}= \left(\f{\partial V}{\partial
u_1},\f{\partial V}{\partial u_2},\f{\partial V}{\partial
u_3}\right)^T,~\f{\partial^2V(u,t)}{\partial
u^2}=(\f{\partial^2V}{\partial u_j \partial u_i})_{i,j=1,2,3}$ and
`T' stands for transposition. Now, we state the following theorem
due to Afanasev \emph{et al}. (1996).
\\
\no \textbf{Theorem 4.1:} \emph{Suppose there exists a function $V(u,t)
\in {C_2}(\Psi)$ satisfying the inequalities
\begin{eqnarray}
&K_1|u|^p \,\leq\, V(u,t)\,\leq\, K_2 |u|^p,\\
&W(u,t) \leq - K_3|u|^p,~K_i > 0,~p>0~(i=1,2,3).
\end{eqnarray}
\no Then the trivial solution of (19) is exponentially p-stable for
$t \geq 0$.}

\par We note that, if $p=2$ in (23) and (24) then the trivial solution
of (19) is exponentially mean-square stable. Furthermore, the
trivial solution of (19) is globally asymptotically stable in
probability (Afanasev \emph{et al}., 1996). From the standard
stability analysis of the stochastic model (14), we state the
following theorem.
\\
\no {\bf Theorem 4.2:} {\it Along with the existence criterion for
$E^*$ (condition (7)-(8)) as stated in section (3.1), if the
following condition holds
\begin{eqnarray}
\gamma_1\,\gamma_2\,>\,
\left(\f{\alpha_3}{r_3}\right)^4\,\left(\f{\sigma_{1}^2\,\sigma_{2}^2\,\sigma_{3}^2}{8\,r_3}\right)
\end{eqnarray}
then the trivial solution of system (14) is asymptotically
mean-square stable.}
\\
\emph{(For proof see Appendix)}

\par We recall that, $\sigma_i$s ($i=1,2,3$) represent the intensities (or rapidity) of the
environmental fluctuations. The above theorem demonstrates that
suitable values of the growth rate of TPP and the intensity of the
toxin-allelopathy parameters determine the stochastic stability.
Hence similar to deterministic environment, toxin allelopathy might
also be a potential candidate for preventing competitive exclusion
among the phytoplankton species in stochastic environment. Recalling
condition (11) for the stability of coexisting equilibrium in the
deterministic model, we find that along with the existence criterion
(in section 3.1), if the product of the strengths of toxin
allelopathy (i.e., $\gamma_1$ and $\gamma_2$) is bigger in magnitude
than
$\max\{\left(\f{\alpha_3}{r_3}\right)^4\,\left(\beta_{12}\,\beta_{21}\right),\,\left(\f{\alpha_3}{r_3}\right)^4\,
\left(\f{\sigma_{1}^2\,\sigma_{2}^2\,\sigma_{3}^2}{8\,r_3}\right)\}$,
then the dynamics of the coexisting competitive-phytoplankton
species is locally stable, both in constant and fluctuating
environment.

\section{Numerical experiments}

\no \bc \emph{Dynamics when no toxic species present} \ec

\no When the toxic phytoplankton is absent in the system, without
any loss of generality, let the two component model have hypothetical parameter values as follows, \\
$r_1=0.6$ day$^{-1}$, $r_2=0.6$ day$^{-1}$, $\alpha_1=0.01$
biomass$^{-1}$ day$^{-1}$, $\alpha_2=0.04$ biomass$^{-1}$
day$^{-1}$. Because species 1 is assumed to be a stronger competitor
than species 2, for the following numerical simulations we take
$\beta_{21}~>~\beta_{12}$. Now the two-component model of the
competitive-phytoplankton species in absence of toxic algae is
simulated. We fix $\beta_{12}=0.02\,$ biomass$^{-1}$ day$^{-1}$, and
vary $\beta_{21}$. A suitable range of the competition coefficient
$\beta_{21}$ is found ($0.021\,\leq\,\beta_{21}~\leq~0.05$) for
which the weak competitor goes to extinction, both in deterministic
and stochastic model (Figure 3(a)-3(b)).
\begin{figure}
\begin{center}
\epsfxsize = 5.5in \epsfysize = 4.5in \epsfbox{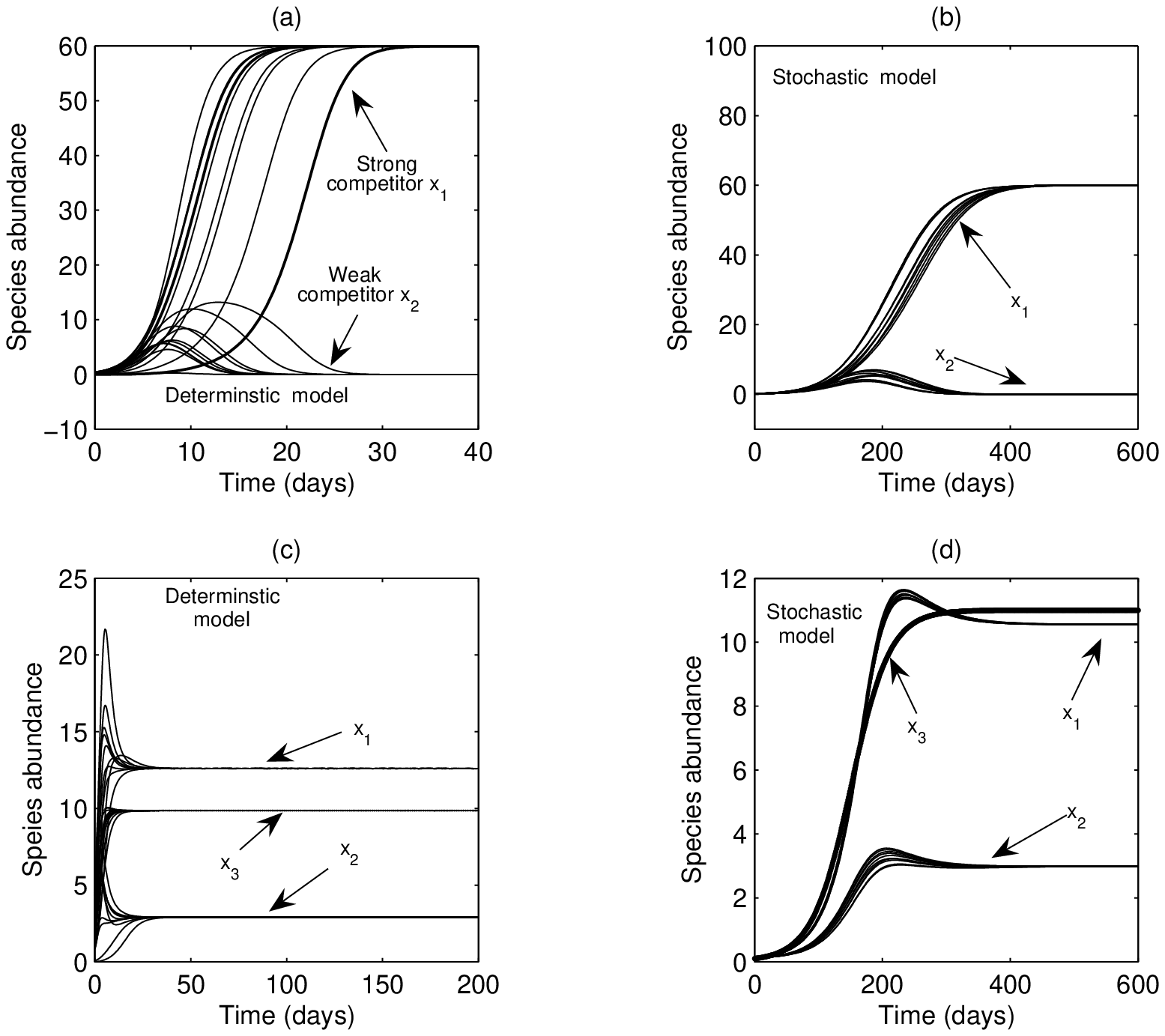}
\end{center}
\no \textbf{Figure 3:} (a) Competitive exclusion of the weak
competitor (species 2) in the absence of TPP for the deterministic
model (1). The fixed parameters are $r_1=0.6\,$day$^{-1}$,
$r_2=0.6\,$ day$^{-1}$, $\alpha_1=0.01\,$biomass$^{-1}\,$day$^{-1}$,
$\alpha_2=0.04\,$biomass$^{-1}\,$ day$^{-1}$, $\beta_{12}=0.02\,$
biomass$^{-1}\,$day$\,^{-1}$; for
$0.021\,\leq\,\beta_{21}~\leq~0.05$, for which species 1 persists
but species 2 goes extinct. (b) Competitive exclusion of the weak
competitor (species 2) in the absence of TPP for the stochastic
model (14), with the parameters fixed as in 3(a) and
$\sigma_1=0.0004$, $\sigma_2=0.0005$. Stable coexistence of the
competitive species in presence of the TPP: (c) deterministic
stability of the model model system (1); all the species coexists in
the same range $0.021\,\leq\,\beta_{21}~\leq~0.05 $, due to
introduction of the TPP with $r_3~=~0.66\,$day$^{-1}$,
$\alpha_3=0.06\,$biomass$^{-1}\,$day$^{-1}$,
$\gamma_1=0.00034\,$biomass$^{-3}\,$day$^{-1}$,
$\gamma_2=0.00006\,$biomass$^{-3}\,$day$^{-1}$ and the other
parameters fixed as in 3(a); (d) stochastic stability of the model
system (14); stable coexistence of all the species even under
stochastic perturbation, with intensity of stochasticity
$\sigma_1=0.00036$, $\sigma_2=0.005$, $\sigma_3=0.00037$.
\end{figure}
\bc \emph{Dynamics when toxic species is included} \ec

\no Now we introduce the toxic algae and simulate the
three-component model with parameters for the two-component model
left unchanged. Suitable values of the toxin-allelopathy parameters
are found for which all the species coexist (Figure 3). The
coexisting equilibrium is stable under deterministic as well as
stochastic set up (Figure 3(c)-(3(d)). A gradual increase in the
intensity of allelopathy show that, a reasonably large range of each
of those allelopathic parameters is obtainable, even beyond the
reported ranges (Sol\'{e} \emph{et al.}, 2005), for which the weak
and the strong species stably coexist with non-zero equilibrium
biomass (Figure 4). In other words, this result
\begin{figure}
\begin{center}
\epsfxsize = 5.5in \epsfysize = 3.5in \epsfbox{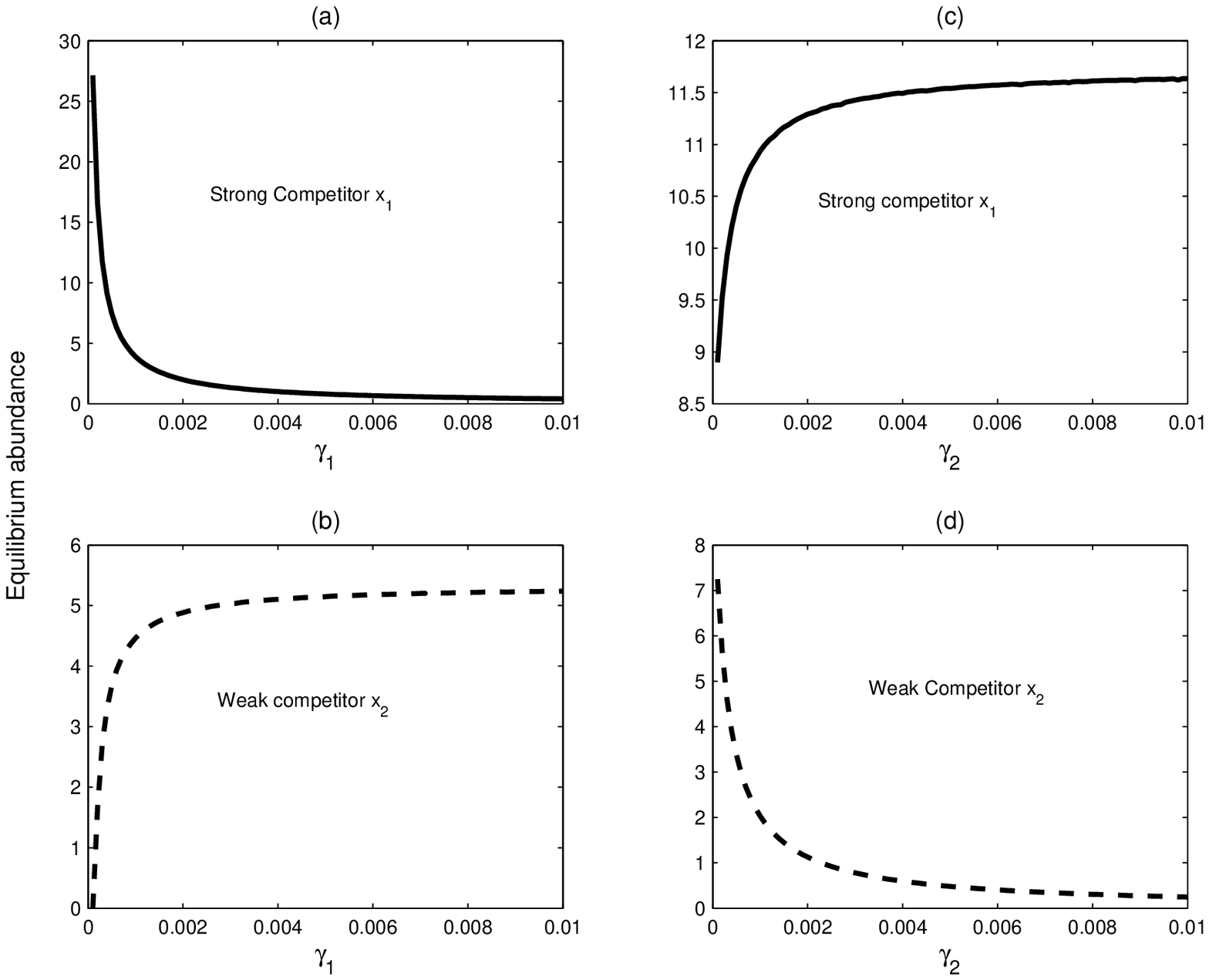}
\end{center}
\no \textbf{Figure 4:} Equilibrium abundance of the strong and weak
competitor with variation in the allelopathic intensity. The
reported range of allelopathic parameter in Sol\'{e} \emph{et al.}
(2005) is $10^{-6}$ to $10^{-5}$. However, stable coexistence is
found for a very large range of $\gamma_1$ and $\gamma_2$. Left
panels (a) $\&$ (b) depicts the variation of the
positive-equilibrium abundance of $x_1$ and $x_2$ for a large range
$0\,\leq\,\gamma_1\,\leq\,0.01$. Right panels (c) $\&$ (d) depict
the same for a large range $0\,\leq\,\gamma_2\,\leq\,0.01$.
\end{figure}
shows that the stable coexistence of the all the species is robust
with respect to the allelopathic effect. Following these arguments,
we suggest that, in the presence of a toxic alga, the possibility
for a competitive exclusion of the weak species of non-toxic
phytoplankton might be overturned.

\bc \emph{Dynamics on consideration of competition coefficient
between TPP and NTP explicitly} \ec

\no As already mentioned, by releasing allelopathic chemicals that
affect the growth of other species, a toxic phytoplankton gains an
advantage in competition. Hence the competing effects of non-toxic
phytoplankton on the toxic phytoplankton is negligible (Hulot $\&$
Huisman, 2004, Roy \emph{et al.} \emph{submitted}), and can be
ignored in modeling the interaction of a non-toxic and toxic algae
(Sol\'{e} \emph{et al.}, 2005). Keeping these observations in mind,
in our three-component model, we have not considered any competition
coefficient between a toxic and a non-toxic phytoplankton. However,
for the the completeness of our study, and for strengthening the
arguments drawn for the stable coexistence of the species, let us
now explicitly introduce the effects of weak competition between
$x_1$-$x_3$ and $x_2$-$x_3$, represented by the coefficients
$\beta_{13}$ and $\beta_{23}$ respectively. Along with the unaltered
forms of the first two equations, the third equation of model system
(1) now takes the following form,
\begin{eqnarray}
\frac{dx_3}{dt} & = & x_3\,\left (
r_3-\alpha_3\,x_3-\beta_{13}\,x_1-\beta_{23}\,x_2\right).
\end{eqnarray}
To examine the effect of the competition coefficients $\beta_{13}$
and $\beta_{23}$, the values of which are very low because of the
presence of toxin allelopathy, we simulate the new form of the model
system (1) with the other parameters kept fixed as in Figure (3).
Again, similar to the previous case, the model shows a stable
coexistence of all the species (Figure 5). Moreover, provided that
the
\begin{figure}
\begin{center}
\epsfxsize = 5.5in \epsfysize = 3.5in \epsfbox{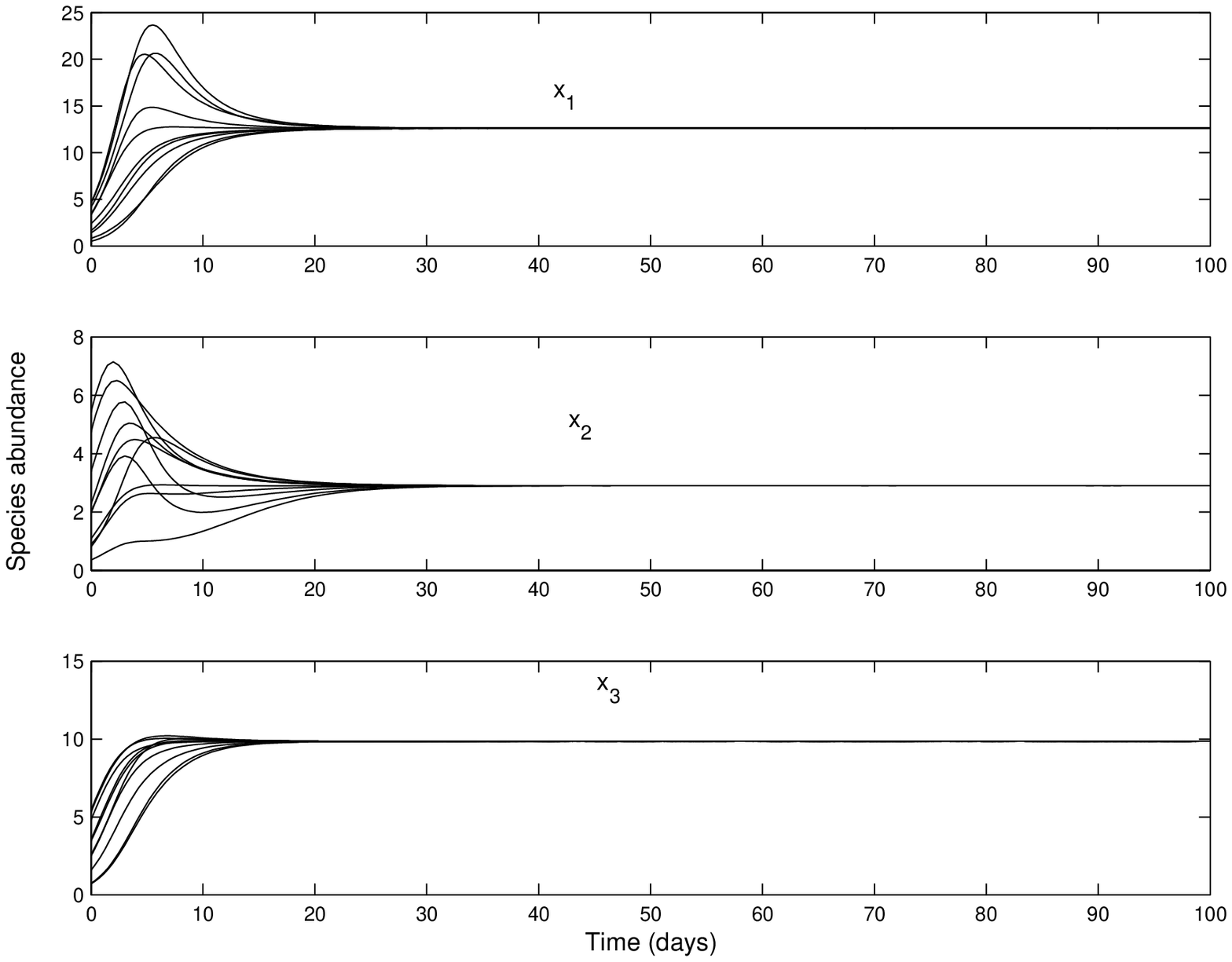}
\end{center}
\no \textbf{Figure 5:} Time series solution of the model system (1),
when the weak competition effect of non-toxic species on the toxic
species is considered explicitly: $\frac{dx_3}{dt} =
x_3\,\left(r_3-\alpha_3\,x_3-\beta_{13}\,x_1-\beta_{23}\,x_2\right)$.
Stable coexistence of all the species for $\beta_{13}=0.005$,
$\beta_{23}=0.002$, and other parameters fixed as before.
\end{figure}
weak-competition coefficients $\beta_{13}$ and $\beta_{23}$ are
bounded within reasonable ranges, and do not attain high values,
positive equilibrium exists and is stable (Figure 6).
\begin{figure}
\begin{center}
\epsfxsize = 5.5in \epsfysize = 3.5in \epsfbox{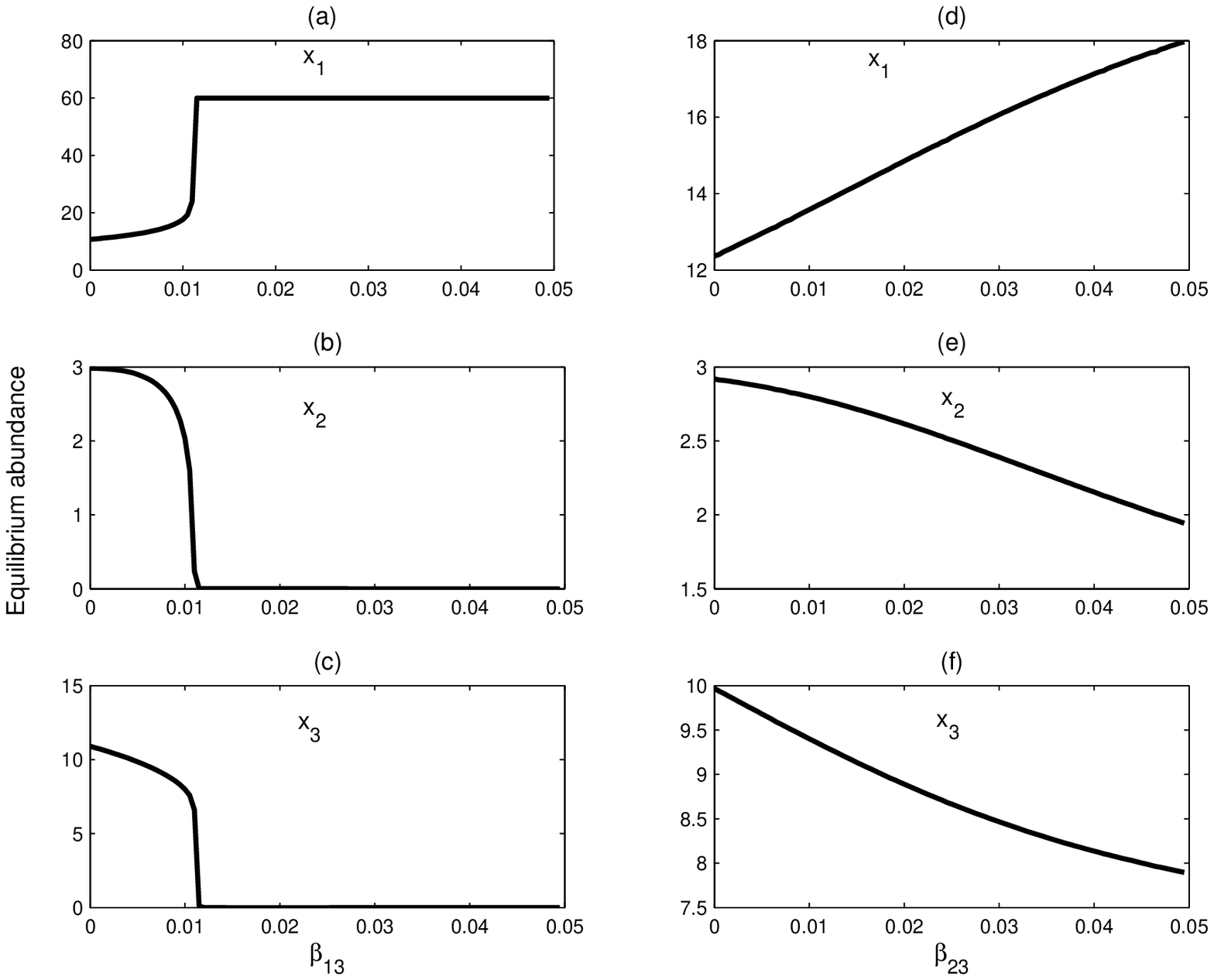}
\end{center}
\no \textbf{Figure 6:} Numerical bounds of the weak competition
coefficients $\beta_{13}$ and $\beta_{23}$ corresponding to other
parameters of the model fixed as in  Figure (1). Left panels (a)-(c)
depict the variation of equilibrium density of $x_1$, $x_2$ and
$x_3$ respectively corresponding to $\beta_{13}$. Right panels
(d)-(f) depicts the same corresponding to $\beta_{23}$. The Figures
depict that the stability of the system tolerates much higher range
of the parameter $\beta_{23}$ than that of $\beta_{13}$; a result
desirable because $x_2$ is considered as the weak competitor that
undergoes competitive exclusion in the absence of the toxic species.
\end{figure}
Because $x_2$ is considered as the weak competitor that undergoes
competitive exclusion in the absence of the toxic species, the
stability of the system tolerates much higher range of the parameter
$\beta_{23}$ than that of $\beta_{13}$ (Figure 6).

\section{Discussion}

\no There is no universally-accepted explanation on how a large
number of species of phytoplankton co-exist on a limited variety of
resources (violating the principle of competitive exclusion). Either
the external factors such as weather or oscillation and chaos
generated by competition among the species were cited for probable
explanations (Hutchinson, 1961; Richerson \emph{et al.}, 1970;
Huisman $\&$ Weissing, 1999). However, in plankton community, the
presence of TPP is remarkable in this context. On an analysis of a
set of field-data that we have collected from the North-West coast
of the Bay of Bengal, here we propose a possible role of
toxin-allelopathy that might be responsible for a stable coexistence
of the competing phytoplankton. Analysis of our field data suggests
that toxic or allelopathic compounds liberated by TPP may be helpful
for reduction of the competition coefficient among phytoplankton
species (Section 2, also claimed by Hulot $\&$ Huisman, 2004). Based
on our field observations and following the study of Sol\'{e}
\emph{et al.} (2005), we have proposed and analyzed a
three-component mathematical model to describe the interactions
among two non-toxic and a toxic phytoplankton. Starting from a
simple two-component Lotka-Volterra competition model representing
competition between two non-toxic phytoplankton, the three-component
model is developed on introducing a third population floor occupied
by a toxic phytoplankton. The analysis in Section 3 demonstrates
that, the strength of toxin-allelopathy determines the coexistence
of the competing-phytoplankton species that would otherwise exhibit
a competitive exclusion of the weak competitor. Restrictions on the
strength of toxin-allelopathy is found that promotes the coexistence
of the phytoplankton species. Moreover, for some ranges of
toxin-allelopathy parameters, the dynamics of the competitors is
stabilized. The conditions for stability, as we have found, are
suitable restrictions on the allelopathic interactions among
non-toxic and toxic phytoplankton species, not driven by external
factors. The dynamics of the competing-phytoplankton species is also
explored in a variable environment. The analysis in Section 4
determines the restrictions on the parameters of toxin-allelopathy
that determine the stability of the coexisting equilibrium under
stochastic fluctuation. These results establish that the growth rate
of toxic phytoplankton and the strength of toxin allelopathy act as
potential parameters for determining the dynamic behaviour of the
competing phytoplankton species, both in a constant and a
fluctuating environment. Finally, for a set of hypothetical
parameters of the model system, numerical simulations have been
performed. Our results show that the possibility of competitive
exclusion among phytoplankton species is overturned because of the
presence of toxin-allelopathy. The overall study suggests that,
although at a species-level interaction toxin-allelopathy due to a
TPP is harmful for the growth of a NTP species, for the competitive
interaction of many NTP species the presence of TPP might be
favourable for the stable coexistence of those species that would
otherwise not coexist. In marine ecosystem where a large number of
phytoplankton species coexist, TPP might be a potential candidate
that, by releasing chemicals, influence on the competitive
interaction among the species, and might promote the survival of the
weak species.

Although the general conclusions drawn from our study follow from
the analysis of the filed samples and that of mathematical models,
we would like to mention some of the limitations of our field study
that could be overcome by a number of complimentary studied. Because
the data that we have used for our analysis is restricted to a field
study, the results of the statistical analysis might associate
factors that are not detectable from a field study without any
laboratory experiments. For instance, although the biomass
coefficient ($x_1/x_2$) of the two non-toxic species has a
significant negative correlation with the abundance of TPP species,
the effect of the abundance of TPP may not be the only cause for
reducing the abundance ratio $x_1/x_2$. There may be several other
causes such as different sensitivity of the analyzed algae to the
toxin or indirect effects such as zooplankton and TPP relationship,
that can be crucial in this context. However, due to lack of
experimental evidences along with our field observations, it is
physically impossible for us to eliminate such effects. Finally, we
suggest that, a number of extensive field studies in multiple
locations would be necessary to establish the implications of our
study in explaining the diversity of phytoplankton in natural
waters.

\no {\bf Acknowledgement:} \emph{The research was supported by a
project fund of the Indian Statistical Institute. The authors are
grateful to the learned referees for their valuable comments on the
previous version of the manuscript.}

\textbf{Appendix}

\textbf{Proof of Theorem (4.2):}
\\
\no Let us consider the following Lyapunov function
\begin{eqnarray} V(u(t),t)=\f{1}{2}\,[\, {u_1}^2 +
w_2\,{u_2}^2 + w_3\,{u_3}^2\,]
\end{eqnarray}
\no where $w_i~(i=1,2,3)$ are real positive-constants to be chosen
suitably. \no It can be easily verified that the inequality (22)
holds for $p=2$. \no Now,
\begin{eqnarray} \f{\partial^2 V}{\partial u^2} = \left [
\matrix{1 & 0 & 0 \cr 0 & w_2 & 0 \cr 0 & 0 & w_3} \right ]
\end{eqnarray}
\no Hence,
\begin{eqnarray} g^T(u(t))\,\f{\partial^2V}{\partial
u^2}\,g(u(t))= \left [ \matrix{{\sigma_1}^2\,{u_1}^2 & 0 & 0 \cr 0
& w_2\,{\sigma_2}^2\,{u_2}^2 & 0 \cr 0 & 0 & w_3\,
{\sigma_3}^2{u_3}^2} \right ],
\end{eqnarray}
\no so, \begin{eqnarray}
\f{1}{2}Tr\,[\,g^T(u(t))\,\f{\partial^2V}{\partial
u^2}\,g(u(t))\,] \,=\, \f{1}{2}\left(\,{\sigma_1}^2{u_1}^2 +
w_2\,{\sigma_2}^2\,{u_2}^2 + w_3\, {\sigma_3}^2\,{u_3}^2\right).
\end{eqnarray}
\no Again
\begin{eqnarray}
f^T ( u(t))\,\f{\partial V(u,t)}{\partial u}\,=\,{\it
x_1^*}\,\left (-{\it \alpha_1}-{\it \gamma_1}\,{{\it
x_3^*}}^{2}\right ){{ \it u_1}}^{2}+{\it x_2^*}\,\left (-{\it
\alpha_2}-{\it \gamma_2}\,{{\it x_3^*}}^{ 2}\right ){\it
w_2}\,{{\it u_2}}^{2}- \nonumber \\  {\it \alpha_3}\,{\it
x_3^*}\,{{\it u_3}}^ {2}{\it w_3}-2\,{\it u_3}\,{{\it
x_2^*}}^{2}{\it \gamma_2}\,{\it x_3^*}\,{\it w_2 }\,{\it
u_2}-2\,{{\it x_1^*}}^{2}{\it \gamma_1}\,{\it x_3}\,{\it
u_3}\,{\it u_1 }-\left ({\it x_2^*}\,{\it \beta_{21}}\,{\it
w_2}+{\it x_1^*}\,{\it \beta_{12}} \right ){\it u_1}\,{\it u_2}
\end{eqnarray}
\no Therefore,
\\
\begin{eqnarray} W(u(t))= -[\f{1}{2}\,{\it w_3}\,\left
(2\,{\it \alpha_3}\,{\it x_3^*}-{{\it \sigma_3}}^{2} \right ){{\it
u_3}}^{2}+ \left (-\f{1}{2}\,{{\it \sigma_1}}^{2}+{\it x_1^*}\,
\left ({\it \alpha_1}+{\it \gamma_1}\,{{\it x_3^*}}^{2}\right
)\right ){{\it u_1}}^{2}+\nonumber
\\
\left ({\it x_2^*}\,{\it w_2}\,\left ({\it \alpha_2}+{\it
\gamma_2}\, {{\it x_3^*}}^{2}\right ) -\f{1}{2}\,{\it w_2}\,{{\it
\sigma_2}}^{2}\right ){{\it u_2}}^{2}+ \left ({\it x_2^*}\,{\it
\beta_{21}}\,{\it w_2}+{\it x_1^*}\,{\it \beta_{12} }\right ){\it
u_1}\,{\it u_2}+\nonumber
\\
2\,{\it u_3}\,{{\it x_2^*}}^{2}{\it \gamma_2}\,{ \it x_3^*}\,{\it
w_2}\,{\it u_2}+2\,{{\it x_1}}^{2}{\it \gamma_1}\,{\it x_3^*}\,{
\it u_3}\,{\it u_1} ].
\end{eqnarray}
\no Let the following conditions hold
\begin{eqnarray}
{\sigma_1}^2\,<\,2\,x_1^*\,\left(\a_1+\g_1\,{x_3^*}^2\right),\,\,
\sigma_2^2\,<\,2\,x_2^*\,\left(\a_2+\g_2\,{x_3^*}^2\right),\,\,
\sigma_3^2\,<\,2\,\a_3\,x_3^*\,
\end{eqnarray}
\no Then $W(u(t))$ can be written in the following form
\begin{eqnarray}
W(u(t))\,=\,-u^T\,Q\,u
\end{eqnarray}
\no where $u\,=\,\left(u_1, u_2, u_3 \right)^T$ and $Q$ is the
following positive definite symmetric-matrix
$$Q\,=\,\left [\begin {array}{ccc} -\f{1}{2}\,{{\it \sigma_1}}^{2}+{\it
x_1^*}\,\left ({ \it \alpha_1}+{\it \gamma_1}\,{{\it
x_3^*}}^{2}\right )& \f{1}{2}\,{\it x_2^*}\,{\it \beta_{21}}\,{\it
w_2}+\f{1}{2}\,{\it x_1^*}\,{\it \beta_{12}}&{{\it x_1^*}}^{2}{\it
\gamma_1}\,{\it x_3^*}\\\noalign{\medskip}\f{1}{2}\,{\it
x_2^*}\,{\it \beta_{21}}\,{ \it w_2}+\f{1}{2}\,{\it x_1^*}\,{\it
\beta_{12}}&{\it x_2^*}\,{\it w_2}\,\left ({\it \alpha_2}+{\it
\gamma_2}\,{{\it x_3^*}}^{2}\right )-\f{1}{2}\,{\it w_2}\,{{\it
\sigma_2}}^{2}&{{\it x_2^*}}^{2}{\it \gamma_2}\,{\it x_3^*}\,{\it
w_2}
\\\noalign{\medskip}{{\it x_1^*}}^{2}{\it \gamma_1}\,{\it x_3^*}&{{\it x_2^*}}^{2
}{\it \gamma_2}\,{\it x_3^*}\,{\it w_2}&\f{1}{2}\,{\it w_3}\,\left
(2\,{\it \alpha_3 }\,{\it x_3^*}-{{\it \sigma_3}}^{2}\right )\end
{array}\right ]
$$
\\
\no Clearly, the eigenvalues of $Q$ namely, $\lambda_1,
\lambda_2,\lambda_3$ are real positive quantities. If $\lambda_m$
denotes the minimum of $\lambda_1, \lambda_2,\lambda_3$, then from
(21) we get
\begin{eqnarray}
W(u,t) \leq -\lambda_m|u(t)|^2
\end{eqnarray}
\no Hence the trivial solution of system (14) is asymptotically
mean-square stable.

\no Sufficient conditions for stability of the model system (1)
under stochastic fluctuation, i.e., system (14), follow
\begin{eqnarray}
{\sigma_1}^2\,<\,2\,x_1^*\,\left(\a_1+\g_1\,{x_3^*}^2\right),~~~~\\
\sigma_2^2\,<\,2\,x_2^*\,\left(\a_2+\g_2\,{x_3^*}^2\right),~~~~\\
\sigma_3^2\,<\,2\,\a_3\,x_3^*\,
\end{eqnarray}
\no Combining the above three conditions and after simplification,
we find the following sufficient condition for stochastic
stability
$$\gamma_1\,\gamma_2\,>\,
\left(\f{\alpha_3}{r_3}\right)^4\,\left(\f{\sigma_{1}^2\,\sigma_{2}^2\,\sigma_{3}^2}{8\,r_3}\right)
$$
\no Hence the theorem.

 \ed